\documentclass[
notitlepage,
nofootinbib,
aps,
pra,
superscriptaddress,
floatfix,
twocolumn,
10pt]{revtex4-2}

\usepackage{bm}
\usepackage{booktabs}

\usepackage{amsmath}
\usepackage{amssymb}
\usepackage{amsfonts}
\usepackage{mathtools}
\usepackage{physics}
\usepackage{microtype}

\usepackage{graphicx}
\usepackage{xcolor}
\colorlet{GREEN}{green}

\usepackage[colorlinks=true,linkcolor=teal,citecolor=teal,urlcolor=teal]{hyperref}
\usepackage{multirow}
\usepackage{orcidlink}
\usepackage[version=4]{mhchem}

\usepackage{tabularx}
\usepackage[normalem]{ulem}
\usepackage[ruled,lined]{algorithm2e}
\SetKw{Continue}{continue}

\usepackage{amsthm}
\newtheoremstyle{sltheorem}
{}                
{}                
{\slshape}        
{}                
{\bfseries}       
{.}               
{ }               
{}                
\theoremstyle{sltheorem}

\theoremstyle{definition}

\newcommand{\Expval}[1]{\langle {#1} \rangle}

\newcommand{\id}{\mathbb{I}}

\newcommand{\Var}[1]{\mathrm{Var}[#1]}

\DeclareMathOperator*{\argmin}{arg\,min}

\begin{document}

\title{Exploiting overcompleteness of Platonic-solid POVMs for shadow estimation}

\author{Dario Melegari\,\orcidlink{0009-0009-3024-5061}}
\email{dario.melegari@algorithmiq.fi}
\affiliation{Dipartimento di Fisica, Universit\`a di Genova, via Dodecaneso 33, I-16146, Genova, Italy}
\affiliation{INFN - Sezione di Genova, via Dodecaneso 33, I-16146, Genova, Italy}
\affiliation{Algorithmiq S.r.l., Via della Chiusa 15, 20123 Milan, Italy}

\author{Stefano Mangini\,\orcidlink{0000-0002-0056-0660}}
\affiliation{Algorithmiq S.r.l., Via della Chiusa 15, 20123 Milan, Italy}
\affiliation{Department of Physics, University of Helsinki, P.O. Box 43, FI-00014 Helsinki, Finland.}

\author{Keijo Korhonen\,\orcidlink{0009-0003-2647-3105}}
\affiliation{Algorithmiq S.r.l., Via della Chiusa 15, 20123 Milan, Italy}
\affiliation{Department of Physics, University of Helsinki, P.O. Box 43, FI-00014 Helsinki, Finland.}

\author{Hetta Vappula\,\orcidlink{0009-0006-6732-6744}}
\affiliation{Algorithmiq S.r.l., Via della Chiusa 15, 20123 Milan, Italy}
\affiliation{Department of Physics, University of Helsinki, P.O. Box 43, FI-00014 Helsinki, Finland.}

\author{Daniel Cavalcanti\,\orcidlink{0000-0002-2704-3049}}
\affiliation{Algorithmiq S.r.l., Via della Chiusa 15, 20123 Milan, Italy}


\begin{abstract}
Accurately estimating expectation values of observables from a finite number of measurement shots is a central challenge in quantum information science. Informationally overcomplete measurements provide a route to reduce estimation variance through optimized classical post-processing. However, the interplay between measurement geometry and dual-frame construction remains largely unexplored. In this work, we study Platonic solid POVMs ---highly symmetric, overcomplete single-qubit measurements whose effects correspond to the vertices of the five Platonic solids on the Bloch sphere--- for the estimation of molecular Hamiltonians. Using $k$-locally optimal dual frames, we show that the geometry of the POVM can have a non-trivial and non-monotonic effect on the estimation variance. We further propose a joint optimization of the POVM orientation and effect weights using a classical proxy state, either a product state or a Matrix Product State (MPS) approximation. We demonstrate that optimized Platonic solid POVMs can outperform standard randomized Pauli measurements, provided the MPS bond dimension is sufficient to faithfully represent the target state. These results reveal a trade-off between classical preprocessing and estimation accuracy, suggesting a practical route to improved observable estimation on near-term quantum hardware.
\end{abstract}

\maketitle


\section{Introduction}

Estimating expectation values of observables with high precision is one of the central tasks in quantum information science. In variational quantum algorithms~\cite{GrimsleyADAPT2019, qAdaptTang2021} and quantum chemistry simulations, one typically needs to evaluate the expectation value of a Hamiltonian $H = \sum_j h_j P_j$, written as a linear combination of Pauli operators $P_j$, where $h_j\in\mathbb R$ are the corresponding coefficients, using a finite number of measurement shots on a quantum processor~\cite{Preskill2018quantumcomputingin}. Since each shot yields a single binary outcome, the estimation is inherently statistical, and reducing the resulting uncertainty is of direct practical importance.

The most straightforward approach is to measure each Pauli term $P_j$ independently in its eigenbasis. While this minimizes the variance for each individual term, it requires a number of distinct measurement settings that scales with the number of terms in the Hamiltonian. A large body of work has addressed this overhead by grouping mutually commuting Pauli operators into jointly measurable sets~\cite{Gokhale2019PauliGrouping, Verteletskyi2020PauliGrouping, Crawford2021PauliGrouping, Zhao2020PauliGrouping}, but these methods still require multiple measurement bases and careful partitioning of the operator set.

A different and increasingly popular strategy is to use \textit{informationally-complete} (IC) measurements, which allow all observables to be estimated simultaneously from a single measurement setting. The classical shadows protocol~\cite{HuangShadows2020}, built on randomized single-qubit measurements, has become a leading method in this direction, with extensions to shallow circuits~\cite{Bertoni2023shallow, Hu2021ShallowShadow, Arienzo2022ShallowShadow} and noisy devices~\cite{ChenRobustShadow2021, KohNoisyShadows2020}. More generally, shadow tomography can be cast in the language of arbitrary measurement frames~\cite{Innocenti2023ShadowTomographyDual, AcharyaShadowTomographyPOVM2021}, whose mathematical backbone is the theory of \textit{dual frames}~\cite{CasazzaFiniteFrames2013, CasazzaIntroFiniteFrames2013}.

An IC measurement is minimal when its effects form a basis for the operator space, and overcomplete (OC) when it contains more effects than this minimum. In the latter case, the effects are linearly dependent, so the reconstruction of an observable from the measurement statistics is not unique: equivalently, the associated dual frame is no longer uniquely determined~\cite{Zhu_2014, Innocenti2023ShadowTomographyDual, KrahmerSparsityDualFrames2013}. This redundancy can be turned into an advantage: by a proper choice of the dual frame, one can reduce the estimation variance for a given observable or set of observables. Recent work has pursued this direction, developing optimal and locally-optimal dual frame constructions~\cite{OptimalProcessingDariano2007, Fischer_2024, Malmi2024EnhancedEstimation, CaprottiDualOptimisation, Mangini2025lowvariance, korhonen2025improvingshadowestimationlocallyoptimal} and demonstrating that substantial reductions in estimation error are achievable through tailored classical post-processing. Another approach to reducing the estimation variance is provided by Locally-Biased Classical Shadows (LBCS)~\cite{hadfield2020measurementsquantumhamiltonianslocallybiased}, which optimizes the local sampling probabilities over the single-qubit Pauli measurement bases $X$, $Y$, and $Z$. To date, LBCS has primarily been investigated within this widely used measurement primitive, whereas its extension to more general, potentially overcomplete POVMs remains comparatively unexplored.

\begin{figure}[t]
    \centering
    \includegraphics[width=0.48\textwidth]{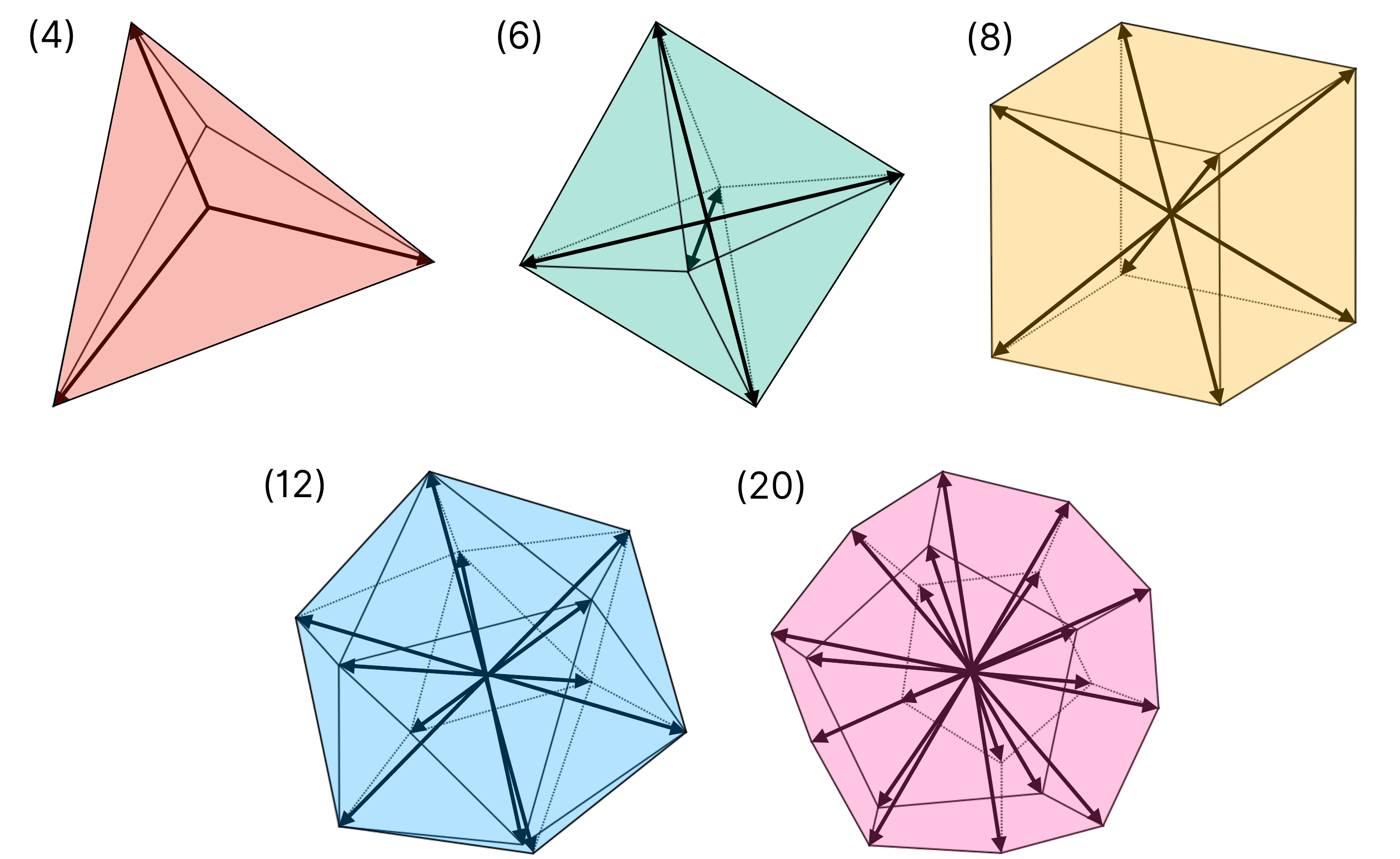}
    \caption{The five Platonic solids. From left to right: the tetrahedron ($4$ effects), the octahedron ($6$), the cube ($8$), the icosahedron ($12$) and the dodecahedron ($20$). The arrows represent the vertices of the solids, which correspond to the effects of the associated POVMs. In the top left of each solid, the number of effects is indicated.}
    \label{fig_platonic_solids}
\end{figure}

In this work, we study a geometrically motivated class of IC measurements whose effects correspond to the vertices of the five Platonic solids ---the tetrahedron, cube, octahedron, dodecahedron, and icosahedron--- placed on the Bloch sphere (see Fig.~\ref{fig_platonic_solids}). 

Platonic-solid POVMs have been considered in various quantum mechanical contexts, from quantum tomography~\cite{Jones1989PhD, Jones_1991_Principles_of_quantum_inference, de_Burgh_2008} to quantum foundations~\cite{Caves2004, Cabello2003, Tavakoli_2020}, and they admit efficient circuit realizations~\cite{Decker2004}. Among these measurements, SIC-POVMs corresponding to the tetrahedron are particularly significant for their optimal information-theoretical properties~\cite{RenesSIC-POVM2004}. The choice of appropriate measurement sets in quantum tomography has been studied in detail~\cite{de_Burgh_2008}, and it was shown that Platonic solid measurements provide good reconstruction performance, with overcomplete measurement sets used to improve the accuracy of tomographic reconstruction. Symmetric measurement configurations are closely related to the notion of highly symmetric POVMs, introduced in~\cite{słomczyński2015highlysymmetricpovmsinformational}, which exhibit extremal entropy and informational power properties. The Platonic POVMs are a natural generalisation of the ubiquitous single-qubit Pauli measurement primitive used in classical shadow techniques, as their symmetry and isotropy make them natural candidates for investigating more general, and potentially more efficient, observable estimators based on highly overcomplete POVMs.

In this work, we systematically evaluate the performance of Platonic-solid POVMs for reducing the variance of observable estimators, specifically combining them with the recently introduced class of locally-optimal dual frames~\cite{korhonen2025improvingshadowestimationlocallyoptimal}. In addition to analysing data via enhanced post-processing methods (the duals), we also study the effect of optimising both the POVM orientation (via single-qubit rotation angles) and the effect weights (the measurement basis probabilities) for the state being measured. We do so by employing either a product-state or Matrix Product State (MPS) approximation as a classical proxy for the true quantum state. We perform numerical experiments on a widely used molecular-Hamiltonian benchmark dataset with up to $16$ qubits~\cite{Hadfield2020Variances}, and demonstrate that optimized Platonic-solid POVMs combined with locally-optimal dual frames can, in some cases, outperform standard approaches based, for example, on Pauli measurements. In particular, we first note an interesting non-monotonic behaviour in estimation performance, in that more complex POVMs are not always associated to more precise estimators. On a similar note, the POVM optimisation on a proxy state is effective provided that the bond dimension of the MPS approximation is sufficient to faithfully represent the target quantum state. These results reveal a resource trade-off: a well-optimized measurement can achieve competitive or superior estimation accuracy while requiring less classical post-processing, at the expense of additional classical pre-processing devoted to measurement optimization.

The paper is organized as follows. Section~\ref{sec_estimating_multiple_observables} reviews the IC measurement framework and the estimation of observables from finite measurement data. Section~\ref{sec_dual_frames} introduces dual frame theory and discusses optimal and locally-optimal dual constructions. Section~\ref{sec_platonic_povms} defines the Platonic-solid POVMs and describes a canonical alignment procedure. Section~\ref{sec_results} presents our numerical results, covering RMSE comparisons across POVMs and the outcome of the orientation optimization. Section~\ref{sec_remarks} summarizes our findings and outlines directions for future work.

\section{Estimating multiple observables simultaneously through informationally complete measurements}\label{sec_estimating_multiple_observables}

A Positive Operator-Valued Measure (POVM) is a central concept in quantum measurement theory~\cite{WatrousQTI2018}. It generalizes the idea of projective measurements by using an arbitrary set of positive operators, which together capture all the statistical information about the outcomes of a quantum experiment. More formally, a quantum measurement with $r$ possible outcomes is described by a set of positive semidefinite operators $\{\Pi_k\}_{k=1}^r$, i.e.\ $\Pi_k \geq 0$ for all $k$, acting on a $d$-dimensional Hilbert space $\mathcal{H}$, that sum to the identity, i.e.\ $\sum_k \Pi_{k} = \id$.

The operators $\Pi_k$ are usually called POVM \textit{effects}, and the probability of the $k$-th outcome when measuring a quantum system in state $\rho$ is $p_k = \mathrm{Tr}[\Pi_k \rho]$. A POVM is called \textit{informationally complete} (IC) if its effects span the full space of linear operators on the Hilbert space $\mathcal{H}$, meaning that any observable $O$ can be decomposed as a weighted sum of the effects:
\begin{equation}
\label{eq:obs_omega}
    O = \sum_{k=1}^{r} \omega_k \Pi_k,
\end{equation}
where $\omega_k \in \mathbb{R}$ are called reconstruction coefficients and depend on the specific observable $O$.

This decomposition immediately yields an expression for the expectation value of $O$ on the state $\rho$:
\begin{equation}
    \Expval{O} = \mathrm{Tr}[O \rho] = \sum_{k} \omega_k \mathrm{Tr}[\Pi_k \rho] = \sum_{k} \omega_{k} p_{k}.
\end{equation}
Thus $\Expval{O}$ can be interpreted as the mean of a random variable associated with the measurement outcomes, where outcome $k$ occurs with probability $p_k$ and is assigned weight $\omega_k$. The reconstruction coefficients $\{\omega_k\}$ therefore provide a direct mapping from the raw measurement statistics of an IC-POVM to the expectation value of any observable.

Following the measurement-frame formulation of Ref.~\cite{Innocenti2023ShadowTomographyDual}, let $o$ denote the single-shot random variable associated with the measurement outcome. When outcome $k$ is observed, with probability $p_k=\mathrm{Tr}[\Pi_k\rho]$, the value $o=\omega_k$ is assigned. Its expectation value and variance are
\begin{align}
    \mathbb{E}[o]
    &= \sum_k p_k\omega_k
    = \Expval{O},\\
    \Var{o}
    &= \sum_k p_k\omega_k^2-\Expval{O}^2.
    \label{eq_single_shot_variance}
\end{align}
Thus, $\Var{o}$ is the variance associated with a single measurement shot.

Let $S$ be the number of measurement shots acquired in an experiment. Given the independent and identically distributed outcomes $o_1,\ldots,o_S$, the expectation value $\Expval{O}$ is estimated using the sample mean:
\begin{equation}
    \bar{o}
    = \frac{1}{S}\sum_{s=1}^{S}o_s
    = \sum_{k=1}^{r} f_k\omega_k,
\end{equation}
where $n_k$ is the number of occurrences of outcome $k$ and $f_k=n_k/S$ is its empirical frequency. The estimator is unbiased,
\begin{equation}
    \mathbb{E}[\bar{o}]=\Expval{O},
\end{equation}
and its variance is
\begin{equation}\label{eq_estimator_variance}
    \Var{\bar{o}} = \frac{\Var{o}}{S} = \frac{1}{S} \left( \sum_k p_k\omega_k^2-\Expval{O}^2 \right).
\end{equation}
An unbiased finite-sample estimator of $\Var{\bar{o}}$ is
\begin{equation}\label{eq_variance_finite_statistics}
    \widehat{\mathrm{Var}}[\bar{o}] = \frac{1}{S-1} \left[ \sum_{k=1}^{r} f_k\omega_k^2 - \left( \sum_{k=1}^{r} f_k\omega_k \right)^2 \right].
\end{equation}
Consequently, by the central limit theorem, for sufficiently large $S$ the sampling distribution of $\bar{o}$ is approximately Gaussian, with mean $\Expval{O}$ and variance $\Var{o}/S$. The statistical fluctuations of the estimator therefore have characteristic magnitude $\sqrt{\Var{o}/S}=\sqrt{\Var{\bar{o}}}$ and decrease as $1/\sqrt{S}$.

To quantify finite-shot performance, consider $R$ independent realizations and let $\bar{o}_j$ denote the estimate obtained in realization $j\in\{1,\ldots,R\}$, with each realization using $S$ sampled shots. A complementary figure of merit, in addition to the variance, is the root mean squared error (RMSE) of the estimator, defined as
\begin{equation}\label{eq_rmse}
    \mathrm{RMSE} = \sqrt{\frac{1}{R} \sum_{j=1}^{R} (\bar{o}_j - \Expval{O})^2}.
\end{equation}
The RMSE is a practical figure of merit because it directly quantifies the expected estimation error in a realistic experimental scenario with finite measurement resources.

For a fixed number of measurement shots $S$, the empirical RMSE computed over $R$ independent realizations converges, as $R\to\infty$, to the square root of the mean-squared error. Because the estimator considered here is unbiased, this quantity coincides with its standard deviation, $\sqrt{\Var{\bar{o}}}=\sqrt{\Var{o}/S}$~\cite{korhonen2025improvingshadowestimationlocallyoptimal}.

\section{Dual frames}\label{sec_dual_frames}

Estimating observables from informationally complete measurements naturally leads to the framework of \textit{dual frames} \cite{Innocenti2023ShadowTomographyDual, CasazzaIntroFiniteFrames2013}. Given an IC-POVM with effects $\{\Pi_k\}$ acting on $\mathcal{H}$, one can construct a corresponding set of operators $\{D_k\}$, called the \emph{dual frame}, which enables the reconstruction of arbitrary observables from the measurement statistics.

More precisely, for any observable $O$ one can write
\begin{equation}
O = \sum_k \mathrm{Tr}[O D_k]\, \Pi_k,
\end{equation}
which implies that the reconstruction coefficients $\omega_k$ are given by
\begin{equation}
\omega_k = \mathrm{Tr}[O D_k].
\end{equation}
Different choices of dual operators lead to different reconstruction coefficients and therefore to different estimation variances~\cite{Innocenti2023ShadowTomographyDual, Zhu_2014, Fischer_2024}.

For \emph{minimal} informationally complete POVMs ($r=d^2$ effects), the dual frame is uniquely determined. However, when the measurement is \emph{overcomplete} ($r>d^2$), the dual frame is no longer unique \cite{Zhu_2014, KrahmerSparsityDualFrames2013}. This freedom can be exploited to reduce the estimation variance through suitable choices of the dual operators.

Several constructions have been proposed in the literature~\cite{Malmi2024EnhancedEstimation, Fischer_2024, CaprottiDualOptimisation, Mangini2025lowvariance, korhonen2025improvingshadowestimationlocallyoptimal}. A commonly used choice is the \emph{canonical dual frame}, which depends only on the POVM itself~\cite{HuangShadows2020}. More generally, if the quantum state $\rho$ were known, one could construct the \emph{optimal dual frame}, which minimizes the estimation variance for the observable being estimated~\cite{Zhu_2014, Innocenti2023ShadowTomographyDual}. Since computing these optimal duals becomes intractable for large systems, practical implementations rely on approximations such as \emph{$k-$locally-optimal} ($k$-LO) dual frames, where the enhanced duals are built from the local reduced density matrices of the state, arising from grouping qubits according to their correlations~\cite{korhonen2025improvingshadowestimationlocallyoptimal}.

For the Platonic-solid POVMs studied here, the use of non-canonical duals is essential, since canonical duals give the same infinite-statistics variance for all uniformly weighted solids considered in our analysis. This follows from the fact that the Platonic solids considered here, excluding the tetrahedron, are at least spherical $3$-designs, while the canonical-dual variance depends only on moments of the measurement directions up to third order. Hence canonical duals are insensitive to the additional overcompleteness of these measurements. The performance differences analyzed below come from exploiting this overcompleteness through non-canonical post-processing, i.e. the $k$-LO duals. Further details are given in Appendix~\ref{sec_canonical_dual_frames}. For this reason, in this work we employ $k$-LO dual frames to process the measurement data. A detailed discussion of dual frames and their explicit construction is provided in Appendix~\ref{app_dual_frames}.

\section{Platonic-solid-based POVMs}\label{sec_platonic_povms}

As discussed in the previous section, exploiting the overcompleteness of a POVM can lead to improved estimation performance for a fixed number of measurement shots. Each Platonic solid defines a single-qubit POVM \cite{Decker2004} by interpreting its vertices as points on the Bloch sphere. The vertices of a Platonic solid depend on its orientation within the Bloch sphere and each orientation corresponds to a distinct POVM. To simplify the discussion, we select specific orientations for the Platonic solids, described in detail below and in Appendix~\ref{app_aligning_platonic_solid_vertices}. 

To implement a POVM corresponding to the same Platonic solid but in a different orientation, it suffices to apply a single-qubit unitary operation to the qubit prior to measurement. All POVMs defined by the vertices of Platonic solids are efficiently implementable and can be realized using a discrete Fourier transform combined with a few additional operations~\cite{Decker2004}.

The five Platonic solids---tetrahedron, octahedron, cube, icosahedron and dodecahedron---have $r = 4$, $6$, $8$, $12$, $20$ vertices respectively, which correspond to the number of POVM effects. Such shapes are highly isotropic and thus would be expected to give good tomographic performance. Indeed, the tetrahedron (also known as Symmetric Informationally Complete, or SIC-POVM) forms a spherical $2-$design~\cite{RenesSIC-POVM2004}, and the cube and octahedron are $3-$designs~\cite{Hardin1996cube}, while the dodecahedron and icosahedron are $5-$designs~\cite{Hardin1996cube} (note that a $t-$design is also a $t'$-design for all $t'<t$)~\cite{de_Burgh_2008}. This motivates us to concentrate on the performance of the Platonic solids and on studying them instead of other possible single qubit POVMs. 

In this work, we exclude the tetrahedron from our analysis for two distinct reasons. First, from an informational standpoint, the tetrahedron-based POVM is IC but not OC, as it contains exactly $r=4$ effects, the minimum required to span the space of operators on a qubit. As a consequence, its dual frame is uniquely determined~\cite{Zhu_2014,Innocenti2023ShadowTomographyDual}, leaving no freedom to optimize classical post-processing. This makes the tetrahedron incompatible with the dual frame optimization strategies that are the focus of this work, which explicitly exploit the redundancy of OC measurements to reduce estimation error. Second, from a practical standpoint, the remaining four solids---octahedron, cube, icosahedron and dodecahedron---admit a NISQ-friendly implementation without ancilla qubits based on randomized projective measurements. In this projective-measurement-simulable implementation, one classically samples an antipodal pair of vertices according to the POVM weights and then measures the qubit projectively along the corresponding Bloch-sphere axis~\cite{Glos2022AdaptivePOVM,Fischer_2024}. More general ancilla-free implementations of POVMs are also possible by exploiting an enlarged qudit space~\cite{Fischer2022AncillaFreePOVM}, but the randomized projective-measurement implementation considered here avoids additional quantum resources and is particularly natural for near-term qubit devices. Moreover, the remaining four Platonic solids are all OC, making them better suited to the dual-frame optimization strategies considered here.

Under the standard Cartesian alignment and with uniform effect weights, the six vertices of the octahedron are the Bloch vectors $\{\pm\hat{x},\pm\hat{y},\pm\hat{z}\}$. The resulting six-effect POVM is therefore exactly the uniform Pauli POVM obtained by choosing the $X$, $Y$, or $Z$ Pauli basis with equal probability and recording the corresponding binary outcome~\cite{HuangShadows2020,hadfield2020measurementsquantumhamiltonianslocallybiased}.

The $n$-qubit POVM is constructed as the tensor product of $n$ single-qubit POVMs, each defined by the same Platonic solid. We label the qubits by $q\in\{1,\ldots,n\}$, and the corresponding Hilbert-space dimension is $d=\dim(\mathcal{H})=2^n$. This results in a total of $r^n$ effects, where $r$ is the number of vertices of the chosen solid. For simplicity, in this study we do not consider mixtures of different solids, i.e., tensor products of single-qubit POVMs defined by different solids.

\subsection{POVM construction with tunable weights}

Given a Platonic solid with $r$ vertices, let $\hat{u}_k \in \mathbb{R}^3$ ($k=1,\dots,r$) be the unit vectors pointing to each vertex on the Bloch sphere, and let $|u_k\rangle$ be the corresponding qubit state. The associated single-qubit POVM effects take the general form
\begin{equation}\label{eq_platonic_povm_general}
    \Pi_k = \nu_k\, |u_k\rangle\!\langle u_k|,
\end{equation}
where $\nu_k \geq 0$ are free \emph{weights} (not to be confused with the outcome probabilities $p_k = \mathrm{Tr}[\Pi_k\rho]$, which also depend on the state $\rho$) subject to the completeness constraint
\begin{equation}\label{eq_weight_normalization_condition}
    \sum_{k=1}^{r} \nu_k |u_k\rangle\!\langle u_k| = \id.
\end{equation}
Equivalently, in the Bloch representation, the weights satisfy $\sum_{k=1}^{r}\nu_k=2$ and $\sum_{k=1}^{r}\nu_k\hat{u}_k=0$. For the centrosymmetric Platonic solids considered below, it is convenient to label the vertices by antipodal pairs $(a,\sigma)$, with $a=1,\ldots,B=r/2$ and $\sigma\in\{+,-\}$. In the optimization procedure, we restrict to weights that are equal within each antipodal pair, $\nu_{a,+}^{(q)}=\nu_{a,-}^{(q)}\equiv \nu_a^{(q)}$, with $\sum_{a=1}^{B}\nu_a^{(q)}=1$ on each qubit $q$. This parametrization preserves the POVM completeness constraint while allowing non-uniform basis probabilities.

The most natural choice is the \emph{uniform} assignment
\begin{equation}\label{eq_uniform_weights}
    \nu_k = \frac{2}{r} \quad \forall\, k.
\end{equation}
As an example, for the octahedron ($r=6$), the six vertices are $\{\pm\hat{x}, \pm\hat{y}, \pm\hat{z}\}$ and Eq.~\eqref{eq_uniform_weights} gives $\nu_k = 1/3$; the resulting POVM is exactly the \emph{uniform Pauli measurement}---choose one of the three Pauli axes uniformly at random and measure in that eigenbasis---the standard single-qubit IC measurement used in classical shadow protocols~\cite{HuangShadows2020}. In the POVM formalism, this classical shadow measurement is represented by the six effects
\begin{align}\label{eq_uniform_pauli_povm}
\begin{split}
    & \Pi_{0} = \frac{1}{3} |{0}\rangle\!\langle{ 0}| \hspace{0.5cm}
    \Pi_{1} = \frac{1}{3} |{ 1}\rangle\!\langle{ 1}| \hspace{0.5cm}
    \Pi_{2} = \frac{1}{3} |{ +}\rangle\!\langle{ +}| \\
    & \Pi_{3} = \frac{1}{3} |{ -}\rangle\!\langle{ -}| \hspace{0.5cm}
    \Pi_{4} = \frac{1}{3} |{ i}\rangle\!\langle{ i}| \hspace{0.5cm}
    \Pi_{5} = \frac{1}{3} |{ -i}\rangle\!\langle{ -i}| .
\end{split}
\end{align}
The same construction applies to every other Platonic solid. The vertices of the Platonic solids are listed in Appendix~\ref{app_platonic_solid_vertices}.

Furthermore, the standard LBCS measurement scheme~\cite{hadfield2020measurementsquantumhamiltonianslocallybiased} is recovered from the six Pauli effects above by replacing the uniform weights with basis-dependent weights. Specifically, assigning $\nu_{+a}=\nu_{-a}=q_a$ for $a\in\{X,Y,Z\}$, with $q_X+q_Y+q_Z=1$, is equivalent to measuring in the Pauli basis $a$ with probability $q_a$~\cite{hadfield2020measurementsquantumhamiltonianslocallybiased}.

\subsection{Choice of the POVM orientation}

In the absence of any prior information about the state to be measured, there is no a priori preferred orientation for a Platonic solid POVM on the Bloch sphere. However, our numerical tests indicate that a convenient default is to align one of the vertices of the solid with the $+z$ axis of the Bloch sphere. This alignment leaves a residual freedom to rotate the solid about the $z$ axis. We fix this freedom by starting from the reference coordinates in Appendix~\ref{app_platonic_solid_vertices} and applying the Rodrigues rotation described in Appendix~\ref{app_aligning_platonic_solid_vertices}, with no additional rotation about $z$. The rationale is twofold: first, it ensures that one POVM effect directly probes the computational basis, which is often the natural measurement basis for quantum hardware; second, in the molecular benchmarks considered here, this alignment yields lower estimation variance than the reference orientations listed in Appendix~\ref{app_platonic_solid_vertices}; see Appendix~\ref{app_aligning_platonic_solid_vertices} for a representative comparison. This canonical alignment is therefore adopted as the default throughout the paper, while in Section~\ref{sec_results} we further optimize the orientation jointly with the effect weights using an MPS proxy state.

By allowing the pair weights $\nu_a^{(q)}$ to deviate from the uniform choice---while still satisfying Eq.~\eqref{eq_weight_normalization_condition}---one obtains a richer family of POVMs whose properties can be tailored to the target state. Different weight distributions change the outcome probabilities $p_k = \mathrm{Tr}[\Pi_k\rho]$ and hence the reconstruction coefficients $\omega_k = \mathrm{Tr}[D_k O]$, directly affecting the estimation variance. This degree of freedom is one of the key levers exploited in Section~\ref{sec_results}, where we jointly optimize the POVM orientation, i.e. the direction of the POVM effects, and the pair weights $\{\bm{\nu}^{(q)}\}_{q=1}^n$ to minimize estimation variance for a target state $\rho$ and Hamiltonian $H$.

\section{Results}\label{sec_results}

The Platonic-solid POVMs are benchmarked on a widely used dataset for assessing the measurement performance of several techniques~\cite{Hadfield2020Variances}. We specifically used a set of molecules from $4$ to $16$ qubits, namely \ce{H2} (4 qubits), \ce{H2} (8 qubits), \ce{LiH} (12 qubits), \ce{BeH2}  (14 qubits), \ce{H2O}  (14 qubits), \ce{NH3} (16 qubits), each obtained via the Jordan--Wigner transformation from their respective fermionic Hamiltonians. For each molecule, we estimate the expectation value of the corresponding Hamiltonian $H$ and quantify the finite-shot performance through the RMSE in Eq.~\eqref{eq_rmse}. For a given POVM, we consider different $k$-LO duals by varying the locality parameter $k$. In the following, the \ce{H2O} results are omitted for brevity, as they have the same qubit count as \ce{BeH2} and exhibit qualitatively similar behaviour.

\subsection{RMSE calculation for different Platonic Solid POVMs}\label{sec_rmse_non_opt}
\begin{figure*}[!ht]
    \centering
    \includegraphics[width=\textwidth]{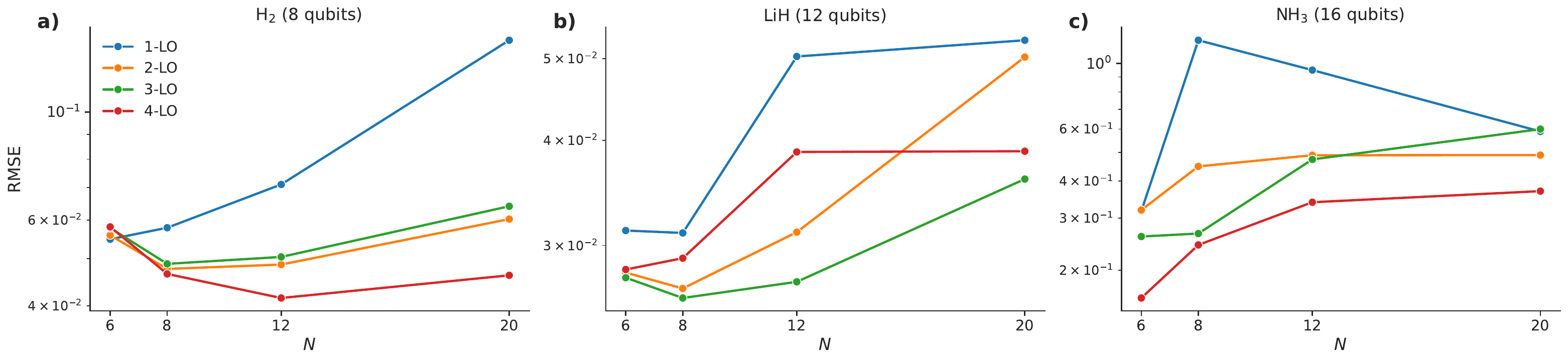}
    \caption{RMSE as a function of the number of effects of the POVM for the \ce{H2} ($8$ qubits), \ce{LiH} ($12$ qubits), and \ce{NH3} ($16$ qubits) molecules. The data show that the best-performing solid depends on both the molecule and the locality parameter of the dual frame: among the plotted settings, \ce{H2} ($8$ qubits) reaches its lowest RMSE with the icosahedron-based POVM and $4$-LO duals, \ce{LiH} ($12$ qubits) with the cube-based POVM and $3$-LO duals, and \ce{NH3} ($16$ qubits) with the octahedron-based POVM and $4$-LO duals. The corresponding numerical values are reported in Tab.~\ref{tab_rmse_non_opt} of Appendix~\ref{app_additional_results_non_optimized_povm}.}
    \label{fig_rmse_non_opt}
\end{figure*}

We begin by benchmarking the finite-shot performance of the different Platonic-solid POVMs through the RMSE of the Hamiltonian estimator. As duals, we considered the $k$-LO dual frame for each POVM and computed the RMSE using Eq.~\eqref{eq_rmse}. The orientation of the POVM is chosen such that one effect is aligned with the $+z$ axis of the Bloch sphere, and the other effects are distributed according to the geometry of the solid, as better explained in App.~\ref{app_aligning_platonic_solid_vertices}. The weights $\{\bm{\nu}^{(q)}\}_{q=1}^{n}$ are taken as uniform, as in Eq.~\eqref{eq_uniform_weights}.

To compare our result with the existing literature, we considered the same experimental setup as in~\cite{korhonen2025improvingshadowestimationlocallyoptimal}, where $R = 10^3$ is the number of independent realizations of the experiment, each using $S=10^3$ measurement shots. The $k$-LO duals used to process the sampled data are obtained from a separate dataset of $S=10^6$ shots.

In Fig.~\ref{fig_rmse_non_opt} we report the RMSE as a function of the number of effects of the POVM for a representative subset of molecules. The results reveal that the optimal POVM geometry is not universal: for small systems, the RMSE displays a non-monotone dependence on the number of effects, indicating the existence of a sweet spot, whereas for the larger examples shown here, increasing the number of effects tends to worsen performance, making POVMs with fewer effects---such as the cube or the octahedron---the preferred choice.

A plausible explanation is a form of proxy-state overfitting induced by the locality constraint. A $k$-LO dual is optimal for the collection of reduced states used in its construction, rather than for the full correlated state. As the POVM becomes more overcomplete, the enlarged reconstruction freedom may allow the dual to fit these local marginals more closely without improving---and potentially while degrading---its performance on correlations extending beyond the chosen groups. The mismatch is expected to become more consequential with increasing system size, because a fixed-$k$ local description represents a progressively smaller fraction of the global correlation structure. This interpretation is consistent with the observation that increasing $k$ often improves performance across POVMs: enlarging the local groups reduces the discrepancy between the surrogate reconstruction problem and the global one. A more detailed investigation of this mechanism is left for future work.

\subsection{Optimising the POVM}

\begin{figure*}[!ht]
    \centering
    \includegraphics[width=\textwidth]{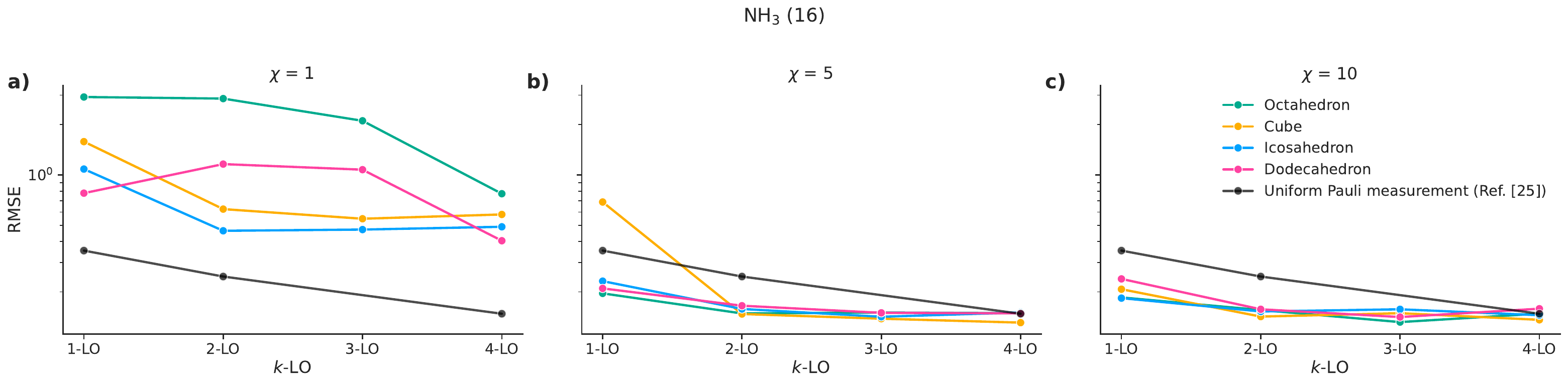}
    \caption{
        RMSE for the \ce{NH3} molecule ($16$ qubits) as a function of the locality parameter $k$ of the $k$-LO duals. The three panels correspond to MPS proxy states with bond dimensions $\chi=1$, $\chi=5$, and $\chi=10$, respectively. The RMSE is computed as in Eq.~\eqref{eq_rmse} using $R=10^3$ independent realizations of the experiment, each with $S=10^3$ measurement shots, and $k$-LO duals obtained from a separate dataset of $S=10^6$ shots.
    }
    \label{fig_nh3}
\end{figure*}

In the preceding analysis, the POVM was kept fixed and only the dual frame was varied. Optimizing the dual is entirely a post-processing operation, since it changes how the measurement data are reconstructed after they have been acquired, without modifying the measurement itself. Here we investigate the complementary possibility of tailoring the POVM to the state and Hamiltonian of interest. This is instead a pre-processing step, because the local orientations and effect weights must be chosen before the quantum experiment is performed. We therefore optimize the measurement on a classically tractable proxy for the target state. The resulting POVM is then implemented on the quantum device, after which the collected data can be processed using the desired dual frame. In our numerical analysis, the optimization minimizes the infinite-statistics variance introduced in Eq.~\eqref{eq_single_shot_variance}, evaluated on the proxy state with fixed $1$-LO duals and written explicitly in Eq.~\eqref{eq_proxy_variance_objective}. The final performance is instead assessed in the finite-statistics regime through the RMSE obtained from samples drawn from the true state and processed with the $k$-LO duals. Thus, the optimization and evaluation stages deliberately use different quantities, with the POVM optimized on the proxy-state variance and the final performance reported through the finite-shot RMSE on the true state.

The simplest classically accessible proxy is a product-state approximation, which neglects interqubit correlations. To incorporate such correlations systematically while retaining classical tractability, we represent the proxy as a matrix product state (MPS)~\cite{Vidal2003EfficientSimulation, Schollwok2011DMRGMPS, PerezGarcia2007MPS}. An MPS encodes a pure $n$-qubit state as a chain of local tensors whose maximum bond dimension $\chi$ controls the amount of entanglement that can be represented. The product-state limit is recovered at $\chi=1$, while increasing $\chi$ yields progressively more expressive approximations and becomes exact, in the worst case, for $\chi=2^{n/2}$. For ground states of gapped local Hamiltonians, the area law for entanglement~\cite{HastingsAreaLaw2007} motivates the use of moderate bond dimensions to capture the correlations relevant to the target Hamiltonian. The bond dimension therefore governs the central accuracy--cost trade-off of this preprocessing strategy: increasing $\chi$ can produce a more faithful proxy and hence a better-informed POVM, at the expense of additional classical computation.

The optimization over the local rotation angles $\{\bm{\theta}^{(q)}\}_{q=1}^n$ and the per-qubit weight vectors $\{\bm{\nu}^{(q)}\}_{q=1}^n$ is performed with the L-BFGS-B algorithm~\cite{byrd1995limited}, a quasi-Newton method that handles box constraints via a limited-memory Hessian approximation. The optimization is run for at most $50$ iterations and is terminated earlier when the relative decrease in the objective function between consecutive iterations falls below $10^{-5}$, which is sufficient to reach convergence (see Appendix~\ref{app_optimization_procedure} for details). The MPS proxy is obtained by truncating the exact ground state~\cite{Hadfield2020Variances} to bond dimension $\chi$, which we vary from $\chi=1$ to $\chi=10$. Once the optimal parameters are found, the finite-shot RMSE is evaluated on the true ground state $\rho$ using the $k$-LO duals.

The RMSE results for the \ce{NH3} molecule are shown in Fig.~\ref{fig_nh3}, with the corresponding numerical values reported in Tab.~\ref{tab_nh3_opt} of Appendix~\ref{app_additional_results_optimized_povm}. The same appendix also reports the results for \ce{H2} ($8$ qubits) and \ce{BeH2} ($14$ qubits), in Tabs.~\ref{tab_h2_opt} and~\ref{tab_beh2_opt}, respectively, together with the corresponding plots in Fig.~\ref{fig_h2_beh2}. The remaining molecules exhibit qualitatively similar behavior.

As shown in Fig.~\ref{fig_nh3}, the optimization procedure reduces the RMSE once the bond dimension $\chi$ of the MPS proxy is sufficiently large. In other words, the optimized POVMs can yield lower RMSE than the reference values of Ref.~\cite{korhonen2025improvingshadowestimationlocallyoptimal}, provided that the proxy state captures the correlations that are relevant for the target Hamiltonian. The corresponding trade-off is the additional classical pre-processing cost required by the optimization procedure.

For $\chi = 1$, which corresponds to a product-state approximation of the true state $\rho$, the optimization procedure does not yield an improvement in RMSE over the uniform Pauli POVM. This is because such a product-state approximation can be far from the true state, so optimizing the proxy variance can lead to a POVM that does not generalize well to the true state $\rho$. As the bond dimension $\chi$ increases, the MPS proxy becomes a better approximation of the true state, and the optimized POVM obtained from the MPS approximation can yield a significant reduction in RMSE compared to the non-optimized POVM. This highlights the importance of using a sufficiently accurate proxy state in the optimization procedure, and the potential benefits of using MPS approximations with moderate bond dimension to capture the relevant correlations in the target state. An analysis of this overfitting behaviour across different values of $\chi$ is reported in Appendix~\ref{app_overfit_analysis}, where we directly compare the variance optimized on the MPS proxy state against the variance evaluated on the true ground state, for the octahedron-based POVM on the \ce{LiH} molecule. This comparison quantifies the gap between what the optimizer targets and what is actually achieved on the true state, and shows that this gap closes as $\chi$ increases.

\subsection{Analysis of the octahedron POVM with fixed orientation}\label{sec_results_optimization_pauli_fixed_orientation}
\begin{figure*}[!ht]
    \centering
    \includegraphics[width=\textwidth]{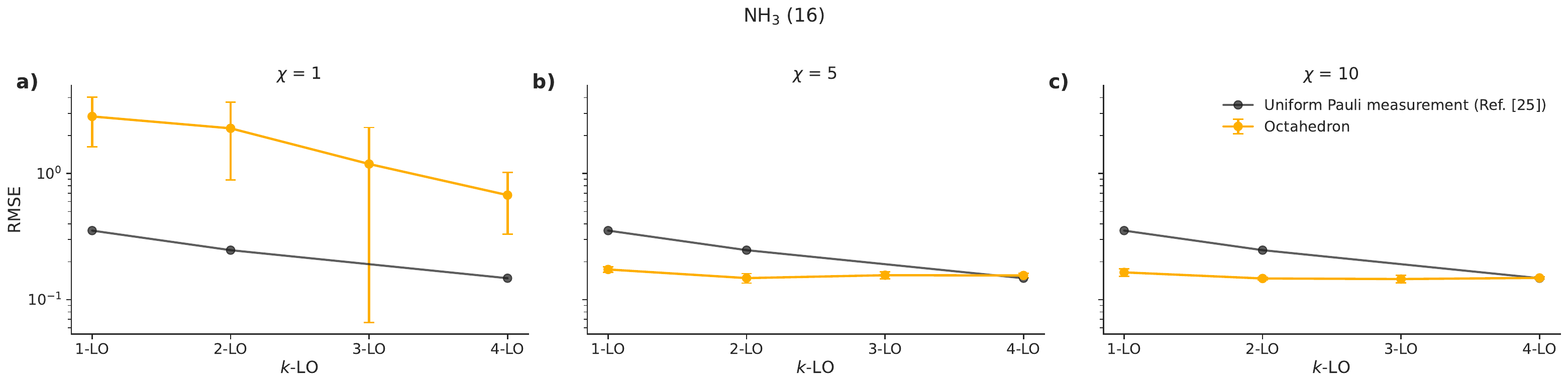}
    \caption{
        RMSE for the \ce{NH3} molecule ($16$ qubits) as a function of the locality parameter $k$ of the $k$-LO duals, for an octahedron-based POVM with fixed orientation. The three panels correspond to MPS proxy states with bond dimensions $\chi=1$, $\chi=5$, and $\chi=10$, used in the optimization procedure. The RMSE (see Eq.~\eqref{eq_rmse}) is computed using $R=10^3$ independent experimental realizations, each with $S=10^3$ measurement shots, while the $k$-LO duals are obtained from a separate dataset of $S=10^6$ shots. Orange markers show the octahedron-based POVM results, with error bars indicating the standard deviation over five independent initializations of the optimization parameters. Black markers show the uniform Pauli measurement reference values from Ref.~\cite{korhonen2025improvingshadowestimationlocallyoptimal}, plotted without error bars because only the tabulated RMSE values are used. The numerical RMSE values are reported in Tab.~\ref{tab_nh3_opt} of Appendix~\ref{app_additional_results_optimized_povm}.
    }
    \label{fig_nh3_errors}
\end{figure*}

As established in Section~\ref{sec_platonic_povms}, the uniform Pauli measurement used in classical shadow protocols~\cite{HuangShadows2020} is exactly the octahedron-based POVM with uniform weights $\nu_k = 1/3$ (see Eq.~\eqref{eq_uniform_pauli_povm}). In most practical applications, this POVM is a natural and convenient choice, as it can be implemented with minimal overhead and does not require any state-dependent optimization. To understand the performance of this POVM, we analyze the RMSE when using the octahedron POVM with fixed orientation (i.e., without any optimization of the rotation angles), by only optimizing the weights $\{\bm{\nu}^{(q)}\}_{q=1}^n$. This analysis allows us to isolate the contribution of the weight optimization from that of the orientation optimization, and to assess the potential benefits of optimizing the weights alone. For the octahedral POVM, these weights have a direct operational interpretation: writing $\nu_{+a}^{(q)}=\nu_{-a}^{(q)}=q_a^{(q)}$ for $a\in\{X,Y,Z\}$, the optimized parameters $q_X^{(q)}$, $q_Y^{(q)}$, and $q_Z^{(q)}$ are precisely the probabilities of measuring qubit $q$ in the $X$, $Y$, and $Z$ bases, respectively. This is closely related to Locally-Biased Classical Shadows (LBCS)~\cite{hadfield2020measurementsquantumhamiltonianslocallybiased}, where the local Pauli-basis probabilities are tailored to a target Hamiltonian using a classically tractable reference state. Here we recover the same state-informed probability-biasing principle within the POVM and dual-frame formalism: the probabilities are optimized on an MPS proxy, the resulting measurement is performed on the true state, and the acquired data are subsequently processed with $k$-LO duals. Fig.~\ref{fig_nh3_errors} shows the RMSE for the \ce{NH3} molecule as a function of the locality parameter $k$ of the $k$-LO duals, for the octahedron POVM with fixed orientation. The three panels correspond to MPS proxy states with bond dimensions $\chi=1$, $\chi=5$, and $\chi=10$. The error bars correspond to the standard deviation across $5$ different runs of the optimization procedure, each with a different random initialization of the weights. The reference line (black) corresponds to the RMSE obtained with the uniform Pauli POVM and the $k$-LO duals, without any optimization of the weights. The numerical RMSE values are reported in Tab.~\ref{tab_nh3_opt} of Appendix~\ref{app_additional_results_optimized_povm}.

Comparing Fig.~\ref{fig_nh3_errors} with the jointly optimized results in Fig.~\ref{fig_nh3} shows that, for the octahedral POVM, optimizing the weights alone captures most of the improvement obtained by optimizing both weights and orientations. Within the variability across random initializations, the additional gain from angular optimization is comparatively modest. This suggests a practical heuristic: retain the canonical Cartesian orientation with one axis aligned with $z$, and optimize only the local Pauli-basis biases. We focus on this restricted setting for two reasons: first, the previous results indicate that it consistently yields near-optimal variance reduction; second, restricting the optimization to the weights substantially reduces the dimensionality of the classical optimization problem compared to jointly optimizing orientations, making it computationally cheaper.

\section{Conclusions}\label{sec_remarks}

In this paper, we investigated the use of Platonic-solid-based POVMs for the estimation of expectation values of molecular Hamiltonians using the framework of dual frames. By leveraging the overcompleteness of such measurement schemes and combining them with the recently proposed locally-optimal dual frames~\cite{korhonen2025improvingshadowestimationlocallyoptimal,Fischer_2024}, we demonstrated that, in some cases, it is possible to reduce the finite-shot estimation error, quantified through the RMSE in Eq.~\eqref{eq_rmse}, beyond what is achievable with simpler POVMs, for example the ubiquitous single-qubit random Pauli measurement primitive used in most of the classical shadow approaches.

We first studied how the overcompleteness of the POVM impacts the statistical performance of the corresponding estimator, observing a non-monotonic behaviour with respect to the number of POVM effects. For some small and intermediate system sizes, there is a sweet spot at an intermediate number of measurement effects for which the estimator reaches its best performance. For the larger systems considered here, instead, increasing the number of effects often worsens the RMSE when the estimator is built using locally optimal duals. We interpret this as a possible consequence of the locality constraint in the $k$-LO dual construction: the additional reconstruction freedom can improve performance on the local surrogate problem used to build the duals, without necessarily improving performance on the full correlated state.

Secondly, we investigated an efficient optimisation scheme by which the Platonic-solid POVMs are tuned to the specific state and observable being measured. This works as a fully classical pre-processing step in which the local orientations and basis probabilities of the POVM are optimised to minimise the variance of the estimator computed on a proxy of the original state, obtained via an MPS approximation whose complexity is controlled by its bond dimension. When the bond dimension is sufficient to capture the relevant features of the state, the POVM optimised on the proxy generalises well to the true state, yielding estimators that can outperform non-optimised approaches.

Motivated by the practical relevance of the octahedral POVM, which coincides with the standard randomized Pauli measurement when uniformly weighted, we also focused on this case separately. We find that optimizing only the local Pauli-basis probabilities captures most of the improvement obtained through the full joint optimization of weights and orientations. This provides a particularly practical variant of our approach, as it reduces the classical optimization cost while retaining most of the observed variance reduction.

Our numerical experiments, performed on a widely used benchmark set of molecular Hamiltonians~\cite{Hadfield2020Variances}, show that the combination of Platonic-solid POVMs with locally optimal dual frames can reduce the statistical error of the estimation compared to simpler measurements, especially when using POVMs tailored to the specific state and observable being measured. It is worth emphasizing, however, that the uniform Pauli POVM, coupled with locally optimal dual frames, remains a strong and practically robust baseline: it requires no classical pre-processing for the measurement itself, is independent of any approximation of the target state, and consistently yields competitive performance across all tested systems. Overall, the use of both non-optimised and optimized Platonic-solid POVMs is therefore best understood as a complementary strategy, offering potential RMSE reduction at the cost of additional classical pre- and post-processing compute.

Several directions are open for future investigation. First, extending the orientation optimization to joint multi-qubit POVMs, rather than product POVMs, could yield further reductions in variance by capturing inter-qubit correlations directly at the measurement level. Second, the integration of the proposed framework with adaptive measurement strategies, where the POVM parameters (probabilities and orientation) are updated on-the-fly based on accumulated measurement data, represents a promising avenue toward fully online estimation protocols. From an optimization perspective, exploring the use of $k$-LO duals during the optimization procedure, rather than fixing to $1$-LO, could potentially lead to even better performance, albeit at the cost of increased computational overhead. Finally, experimental validation of the proposed methods on near-term quantum hardware would be an important step toward assessing their practical utility in real-world quantum applications.

\section*{Acknowledgements}

D.M. acknowledges financial support from INFN. We thank Chemistry teams at Algorithmiq for molecule Hamiltonians used for our benchmarks.


\bibliographystyle{apsrev4-2}
\bibliography{bibliography}

\appendix

\section{Dual frames}\label{app_dual_frames}
A common method for constructing unbiased estimators of observables from IC measurement data relies on the framework of the so-called \textit{dual frames} \cite{Innocenti2023ShadowTomographyDual, DArianoICMeasurements2004, OptimalProcessingDariano2007}. Consider an IC-POVM with effects $\{\Pi_k\}$. A collection of operators $\{D_k\}$ is said to form a \textit{dual frame} associated with $\{\Pi_k\}$ if, for any operator $O$, the following reconstruction identities are satisfied:
\begin{equation}
\label{eq_dual_frame_decomposition}
O = \sum_k \mathrm{Tr}[O D_k]\, \Pi_k
  = \sum_k \mathrm{Tr}[O \Pi_k]\, D_k.
\end{equation}
These relations imply that the coefficients used to reconstruct $O$, as introduced in Eq.~\eqref{eq:obs_omega}, can be written as
\begin{equation}\label{eq_omega_k}
    \omega_k = \mathrm{Tr}[O D_k],
\end{equation} 
where the operators $\{D_k\}$ serve as the dual frame (or dual effects) corresponding to the POVM $\{\Pi_k\}$.

When an IC-POVM consists of exactly $r= d^2$ measurement outcomes, it is referred to as a minimal IC-POVM, in which case the associated dual effects are uniquely determined. Here $d = \dim(\mathcal{H})$ denotes the Hilbert space dimension. By contrast, if the measurement includes more than $r> d^2$ effects, the POVM is informationally overcomplete (OC). In this situation, some of the effects are linearly dependent on others, and consequently the dual frame is no longer unique~\cite{Innocenti2023ShadowTomographyDual, DArianoICMeasurements2004, Zhu_2014}. This opens the possibility of choosing different dual frames, which can lead to different estimation variances for the same observable when using the corresponding estimators.

\subsection{Canonical dual frames}\label{sec_canonical_dual_frames}
One specific and widely used choice of dual frame is provided by the
canonical duals, which are defined as~\cite{Innocenti2023ShadowTomographyDual, Zhu_2014, HuangShadows2020}
\begin{equation}
D_k = \frac{1}{\mathrm{Tr}[\Pi_k]}\, F^{-1}(\Pi_k),
\label{eq:canonical_duals}
\end{equation}
where $F$ denotes the frame operator, a linear map defined as 
\begin{equation}
    F(\cdot) = \sum_{k} \frac{1}{\mathrm{Tr}[\Pi_k]} \mathrm{Tr}[\cdot \, \Pi_k] \, \Pi_k.
\end{equation}
Note that this choice of dual frame corresponds to using the usual ``classical shadow" of  the state in the shadow tomography terminology.

Importantly, the numerical results of Section~\ref{sec_results} do not include a comparison across Platonic solid POVMs under the canonical reconstruction. For the uniformly weighted, centrosymmetric Platonic POVMs considered in the numerical analysis, the canonical reconstruction gives the same infinite-statistics variance.

This can be understood from the fact that the octahedron, cube, icosahedron, and dodecahedron form spherical designs of order at least three~\cite{de_Burgh_2008}. Since the canonical-dual estimation variance depends on moments up to third order in the measurement directions~\cite{Innocenti2023ShadowTomographyDual, HuangShadows2020, Mele2024introductiontohaar}, these \textit{uniformly} weighted POVMs produce the same canonical-dual variance, independently of the number of effects. 

To summarise, within the uniformly weighted centrosymmetric family considered here, Platonic solid POVMs are equivalent when paired with the canonical dual in the infinite-statistics regime. Finite-shot estimates can still differ because of statistical fluctuations. This underlines the role of non-canonical duals when the POVM is overcomplete.

\subsection{Optimal dual frames}\label{sec_optimal_dual_frames}

As discussed in Section~\ref{sec_dual_frames}, whenever the number of POVM effects $r$ is such that $r > d^2$, the POVM is said to be \textit{overcomplete} (OC), and its dual frame is not uniquely defined~\cite{Zhu_2014, KrahmerSparsityDualFrames2013}. Furthermore, different choices of dual operators yield estimators with different statistical properties, due to the dependence of the variance on the reconstruction coefficients $\omega_k = \mathrm{Tr}[D_k O]$.

If one has access to the quantum state $\rho$ to be measured, then there is an optimal choice of dual operators that minimizes the state-dependent quadratic term entering the estimation variance. These optimal duals are defined as~\cite{Innocenti2023ShadowTomographyDual, Zhu_2014}
\begin{align}
    D_k^{\mathrm{opt}} &= \frac{1}{\mathrm{Tr}[\Pi_k \rho]}\,F_{\mathrm{opt}}^{-1}(\Pi_k), \qquad \\
    F_{\mathrm{opt}}(\cdot) &= \sum_{k=1}^{r} \frac{1}{\mathrm{Tr}[\Pi_k \rho]}\,\mathrm{Tr}[\,\cdot\,\Pi_k]\,\Pi_k,
    \label{eq_optimal_duals}
\end{align}
which explicitly depend on the outcome probabilities $p_k = \mathrm{Tr}[\Pi_k \rho]$. These duals can be shown to minimize both the expected mean-squared error for state reconstruction and, more importantly for our purposes, the variance contribution associated with any fixed observable~\cite{Fischer_2024}.

Importantly, the same state-dependent dual construction applies to all observables: for a fixed IC measurement, the optimal post-processing rule is determined by the state $\rho$, while the observable enters only through the coefficients $\omega_k=\mathrm{Tr}[O D_k]$. The main limitation of this approach is that $\rho$ is generally unknown prior to measurement~\cite{korhonen2025improvingshadowestimationlocallyoptimal}, and most importantly, the frame operator $F_{\mathrm{opt}}$ has dimension $d^2 \times d^2$, making its inversion computationally intractable for large systems. As a result, the optimal duals cannot be, in general, directly implemented. One way to overcome this issue is to introduce a locality constraint, leading to the concept of $k$-LO duals~\cite{Fischer_2024,korhonen2025improvingshadowestimationlocallyoptimal}, where the optimisation is performed independently on small groups of qubits. This approach is described in detail in the following section.

\subsection{Locally-optimal dual frames}\label{sec_locally_optimal_dual_frames}

The optimal duals~\eqref{eq_optimal_duals} require both full knowledge of $\rho$ and the inversion of an exponentially large frame operator, making them intractable beyond a few qubits. A practical workaround, introduced in Ref.~\cite{korhonen2025improvingshadowestimationlocallyoptimal}, replaces the global optimisation with a collection of small, independent ones: one partitions the $n$ qubits into disjoint groups of size at most $k$, reconstructs the reduced state on each group from the measurement data, and builds the optimal duals~\eqref{eq_optimal_duals} separately within each subspace. The resulting objects are called $k$-locally optimal ($k$-LO) duals.

In practice, let $\mathcal{G} = (G_1, \dots, G_L)$ be a partition of the qubits with $|G_\ell| \leq k$. For each group $G_\ell$, let $\boldsymbol{k}_{G_\ell}=(k_q)_{q\in G_\ell}$ denote the corresponding local outcome multi-index. One obtains the reduced density matrix $\rho_{G_\ell} = \mathrm{Tr}_{\bar{G}_\ell}[\rho]$ and constructs the local frame operator
\begin{equation}
    F_{G_\ell}(\cdot) = \sum_{\boldsymbol{k}_{G_\ell}} \frac{1}{\mathrm{Tr}[\Pi^{G_\ell}_{\boldsymbol{k}_{G_\ell}} \rho_{G_\ell}]} \, \mathrm{Tr}[\Pi^{G_\ell}_{\boldsymbol{k}_{G_\ell}} \cdot] ~\Pi^{G_\ell}_{\boldsymbol{k}_{G_\ell}}\,,
    \label{eq_klo_frame}
\end{equation}
where the sum runs over all local outcomes $\boldsymbol{k}_{G_\ell}$. The locally-optimal duals on this group $D^{G_\ell}_{\boldsymbol{k}_{G_\ell}}$ are built with Eq.~\eqref{eq_optimal_duals}, and the full $n$-qubit dual is assembled as the tensor product over the different groups
\begin{equation}
    D_{\boldsymbol{k}} = \bigotimes_{G \in \mathcal{G}} D^{G}_{\boldsymbol{k}_G}.
    \label{eq_klo_global}
\end{equation}

The grouping is guided by the pairwise classical mutual information of the measurement outcomes: a greedy algorithm repeatedly selects the most correlated pair among the remaining qubits, grows the group up to size $k$, and then starts a new group~\cite{korhonen2025improvingshadowestimationlocallyoptimal}. This ensures that the strongest correlations in the state are preserved within the blocks where the duals can exploit them.

One can show that the $k$-LO duals are exactly the optimal duals for the product state $\rho_{\mathcal{G}} = \bigotimes_\ell \rho_{G_\ell}$, which approximates $\rho$ by retaining intra-group correlations while discarding those between different groups. The closer $\rho$ is to this block-product form, the closer the $k$-LO duals are to the true optimum. In the two extreme cases, $k=1$ reduces to a product of single-qubit optimal duals, while $k=n$ recovers the globally optimal construction, as given in Eq.~\eqref{eq_optimal_duals}. Since each local frame operator has dimension at most $4^{k}$, its inversion remains efficient for moderate values of $k$ (typically $k \lessapprox  4$ in practice).

In our simulations, we used the \texttt{KaHyPar} grouping method~\cite{korhonen2025improvingshadowestimationlocallyoptimal, schlag2021highqualityhypergraphpartitioning}, a state-of-the-art hypergraph partitioning algorithm.

\section{Details on the orientation of the Platonic solid POVMs}

\subsection{Platonic Solid Vertices}\label{app_platonic_solid_vertices}

Each Platonic solid inscribed in the unit sphere defines a set of vertices on the Bloch sphere that can be used to construct a symmetric POVM.
Tab.~\ref{tab_platonic_vertices} lists the vertices of each solid, grouped up to antipodal pairs.
Since all Platonic solids except the tetrahedron are centrosymmetric (i.e., symmetric under inversion through the origin), for those solids it suffices to list one representative vertex per antipodal pair, the opposite vertex being simply its negation.
The tetrahedron is the sole exception: its four vertices are listed in full, as no two of them are antipodal.
The golden ratio $\phi = (1+\sqrt{5})/2$ appears in the coordinates of the icosahedron and dodecahedron; their vertices are given in unnormalized form and must be divided by their respective norms ($\sqrt{1+\phi^2}$ and $\sqrt{3}$) to lie on the unit sphere.
The coordinates listed in Tab.~\ref{tab_platonic_vertices} are a standard representation of the Platonic-solid vertices in the literature~\cite{de_Burgh_2008, Decker2004}, and we adopt them here as our reference orientation before applying the alignment rotation described in the next section.

\begin{table*}[t]
\centering
\begin{tabular}{ccc}
\toprule
Solid & Vertices (one per antipodal pair) & Norm \\ 
\midrule

tetrahedron & 
$(1,1,1)$, 
$(-1,-1,1)$, 
$(-1,1,-1)$, 
$(1,-1,-1)$ & $\sqrt{3}$ \\[2mm]

cube & 
$(1,1,1)$,
$(1,1,-1)$,
$(1,-1,1)$,
$(-1,1,1)$ & $\sqrt{3}$ \\[2mm]

octahedron & 
$(1,0,0)$,
$(0,1,0)$,
$(0,0,1)$ & $1$ \\[2mm]

icosahedron & 
$(0,1,\phi)$,
$(0,-1,\phi)$,
$(1,\phi,0)$,
$(-1,\phi,0)$,
$(\phi,0,1)$,
$(\phi,0,-1)$ & $\sqrt{1+\phi^2}$ \\[2mm]

dodecahedron & 
$(1,1,1)$,
$(1,1,-1)$,
$(1,-1,1)$,
$(-1,1,1)$, \\
& $(0,1/\phi,\phi)$,
$(0,-1/\phi,\phi)$,
$(1/\phi,\phi,0)$,
$(-1/\phi,\phi,0)$,
$(\phi,0,1/\phi)$,
$(\phi,0,-1/\phi)$ & $\sqrt{3}$ \\[2mm]

\bottomrule
\end{tabular}

\vspace{2mm}
\caption{Vertices of the Platonic solids inscribed in the unit sphere, listed up to antipodal pairs. The tetrahedron is the only solid that is not centrosymmetric, so all four of its vertices are listed explicitly. For the remaining solids, only one vertex per antipodal pair is given, the opposite vertex being its negation. Each vertex must be divided by the norm listed in the rightmost column to obtain the corresponding unit vector on the Bloch sphere. The golden ratio $\phi = (1+\sqrt{5})/2$ appears in the coordinates of the icosahedron and dodecahedron.}
\label{tab_platonic_vertices}
\end{table*}

\subsection{Aligning Platonic Solid Vertices Along the \texorpdfstring{$z$}{z}-axis}\label{app_aligning_platonic_solid_vertices}

\begin{figure}[ht]
    \centering
    \includegraphics[width=0.48\textwidth]{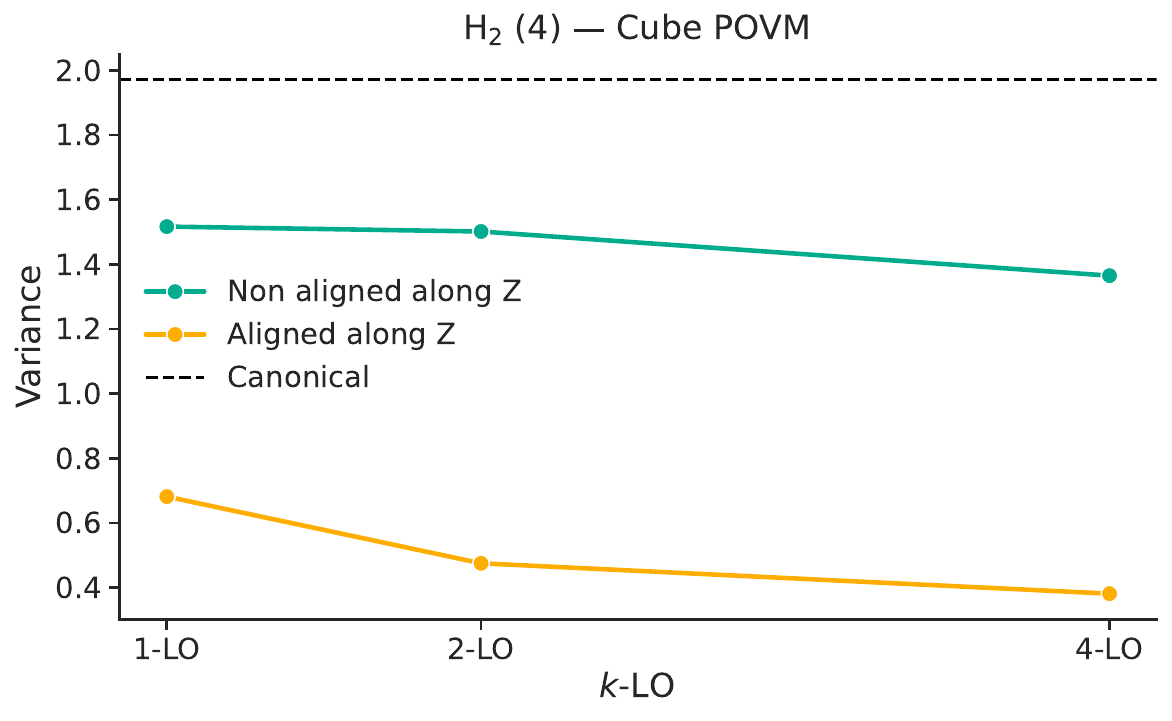}
    \caption{Estimation variance as a function of the $k$-LO duals for the \ce{H2} (4 qubits) molecule and the uniform cube POVM. The variance is computed using Eq.~\eqref{eq_variance_finite_statistics}. The orange solid line corresponds to the variance obtained by aligning one of the vertices of the cube with the $+z$ axis, while the green line corresponds to the variance obtained by choosing the vertices as in Tab.~\ref{tab_platonic_vertices}. The dashed black line corresponds to the variance obtained with the canonical duals. In this example, the aligned orientation yields a smaller variance than the reference orientation. This motivates the alignment procedure described in the text.}
    \label{fig_variance_non_opt}
\end{figure}

Without prior information about the state being measured, there is in principle no preferred orientation for the POVM on the Bloch sphere. Nevertheless, as shown in Fig.~\ref{fig_variance_non_opt}, our numerical results indicate that aligning one vertex of the Platonic solid with the $+z$ axis provides a consistently favourable default orientation. This choice makes one of the POVM effects aligned with a computational-basis state, which is a natural reference direction for the molecular Hamiltonians considered here after the Jordan--Wigner mapping. Similar behaviour is observed for the other molecules and solids in our benchmark set. Below, we describe the rotation procedure used to impose this alignment.

Let $\{\hat{\bm v}_i\}_{i=0}^{r-1} \subset \mathbb{R}^3$ be the set of unit vectors representing the vertices of a Platonic solid on the Bloch sphere, where $r$ is the number of vertices (i.e. the effects) of the POVM. The goal is to rotate all vectors so that one of the vertices $\hat{\bm v}_0$ aligns with the $+z$ axis while preserving the relative positions of the other vertices. Let $\hat{\bm z}=(0,0,1)^T$ denote the unit vector along the positive $z$ axis. We compute the rotation axis $\bm u$ from the cross product between the selected vertex and the $z$ axis,
\begin{equation}
\bm u = \hat{\bm v}_0 \times \hat{\bm z}.
\end{equation}
The rotation angle $\theta$ is obtained from the dot product
\begin{equation}
\cos\theta = \hat{\bm v}_0 \cdot \hat{\bm z}, \quad 
\sin\theta = \|\bm u\|.
\end{equation}
Then, normalize the rotation axis
\begin{equation}
\hat{\bm u} = \frac{\bm u}{\|\bm u\|}.
\end{equation}
We then construct the Rodrigues rotation matrix $R_{\mathrm{rot}} \in \mathbb{R}^{3 \times 3}$ around $\hat{\bm u} = (u_x, u_y, u_z)^T$ by angle $\theta$,
\begin{equation}
    R_{\mathrm{rot}} = I_3 + (\sin\theta) K + (1 - \cos\theta) K^2,
\end{equation}
where $I_3$ denotes the $3\times3$ identity matrix and $K$ is the skew-symmetric matrix associated with the rotation axis $\hat{\bm u}$,
\begin{equation}
    K = 
    \begin{bmatrix}
    0 & -u_z & u_y \\
    u_z & 0 & -u_x \\
    -u_y & u_x & 0
    \end{bmatrix}.
\end{equation}
Finally, we apply the rotation to all vertices,
\begin{equation}
    \hat{\bm v}_i' = R_{\mathrm{rot}} \hat{\bm v}_i, \quad i = 0, \dots, r-1.
\end{equation}
After this transformation, the first vector $\hat{\bm v}_0'$ lies exactly along the $+z$ axis, while all other vertices maintain their relative arrangement on the Bloch sphere. 

The transformation is a rigid rotation, so it preserves the symmetry and isotropy properties of the original Platonic solid.

\section{Details on the optimization procedure}\label{app_optimization_procedure}

As described in the main text, we optimize the POVM parameters by minimizing the infinite-statistics variance evaluated on an MPS proxy $\tilde\rho$ of the target state $\rho$~\cite{Vidal2003EfficientSimulation,Schollwok2011DMRGMPS,PerezGarcia2007MPS}. Let $r$ be the number of local POVM effects and $B=r/2$ the number of antipodal vertex pairs. We label a local outcome by $(a,\sigma)$, where $a\in\{1,\ldots,B\}$ identifies an antipodal pair and $\sigma\in\{+,-\}$ identifies one of its two vertices. If $P_{a,\sigma}$ is the corresponding rank-one projector in the canonical orientation, let $U^{(q)}(\bm\theta^{(q)})$ denote the local unitary parameterized by the angle vector $\bm\theta^{(q)}$. We use $\bm\xi$ to denote collectively all POVM parameters, specified explicitly below. The effect acting on qubit $q$ is
\begin{equation}
    \Pi_{a,\sigma}^{(q)}(\bm\xi)=\nu_a^{(q)}U^{(q)}(\bm\theta^{(q)})P_{a,\sigma}U^{(q)\dagger}(\bm\theta^{(q)}),
\end{equation}
where $\bm\nu^{(q)}=(\nu_1^{(q)},\ldots,\nu_B^{(q)})$ satisfies $\nu_a^{(q)}\geq0$ and $\sum_a\nu_a^{(q)}=1$. Since $P_{a,+}+P_{a,-}=\id$, these conditions imply $\sum_{a,\sigma}\Pi_{a,\sigma}^{(q)}=\id$. For the joint outcome $\boldsymbol{k}=((a_1,\sigma_1),\ldots,(a_n,\sigma_n))$, the product-POVM effect is
\begin{equation}
    \Pi_{\boldsymbol{k}}(\bm\xi)=\bigotimes_{q=1}^{n}\Pi_{a_q,\sigma_q}^{(q)}(\bm\xi).
\end{equation}
We parameterize the basis probabilities using logit vectors $\bm x^{(q)}\in\mathbb R^B$ and fix the minimum probability to $\nu_{\min}=0.05$,
\begin{equation}
    \nu_a^{(q)}(\bm x^{(q)})=\nu_{\min}+\left(1-B\nu_{\min}\right)\frac{e^{x_a^{(q)}}}{\sum_{b=1}^{B}e^{x_b^{(q)}}}.
\end{equation}
The local unitary $U^{(q)}$ is parameterized by three Euler angles $\bm\theta^{(q)}\in[0,2\pi]^3$. The logits are restricted to $\bm x^{(q)}\in[-5,5]^B$ to prevent the optimized probabilities from approaching the lower bound too closely and to improve numerical stability. We collect all rotation angles and logits in $\bm\xi=(\bm\theta^{(1)},\ldots,\bm\theta^{(n)},\bm x^{(1)},\ldots,\bm x^{(n)})$. These intervals define the box constraints used by the optimizer. For a given $\bm\xi$, we define
\begin{equation}
    p_{\boldsymbol{k}}(\boldsymbol{\xi}\, |\, \tilde{\rho})  =\mathrm{Tr}[\tilde\rho\,\Pi_{\boldsymbol{k}}(\bm\xi)],\qquad \omega_{\boldsymbol{k}}(\bm\xi)=\mathrm{Tr}[O D_{\boldsymbol{k}}(\bm\xi)],
\end{equation}
where $D_{\boldsymbol{k}}(\bm\xi)$ is constructed using the $1$-LO prescription throughout the optimization. The objective is therefore
\begin{align}
\label{eq_proxy_variance_objective}
    \Var{O}(\bm{\xi}; \tilde{\rho})
    &=
    \sum_{\boldsymbol{k}}
    p_{\boldsymbol{k}}(\boldsymbol{\xi}\, |\, \tilde{\rho})\,
    \omega_{\boldsymbol{k}}^2(\bm\xi)
    \notag\\
    &\quad
    -\left[
    \sum_{\boldsymbol{k}}
    p_{\boldsymbol{k}}(\boldsymbol{\xi}\, |\, \tilde{\rho})\,
    \omega_{\boldsymbol{k}}(\bm\xi)
    \right]^2,
\end{align}
so that the optimization problem is
\begin{equation}\label{eq_opt_problem_new}
    \bm{\xi}_{\text{opt}} = \argmin_{\bm{\xi}} \Var{O}(\bm{\xi}; \tilde{\rho}).
\end{equation}

We solve this box-constrained problem using \texttt{L-BFGS-B}~\cite{byrd1995limited}. Let $\mathcal U[a,b]$ denote the uniform distribution on $[a,b]$. For each restart, the angles and logits are initialized independently according to $\theta_{j,0}^{(q)}\sim\mathcal U[0,2\pi]$, for $j=1,2,3$, and $x_{a,0}^{(q)}\sim\mathcal U[-1,1]$, respectively. Each run is limited to at most $50$ iterations and is terminated earlier when the relative change in the objective between consecutive iterations falls below $10^{-5}$; this threshold defines the function tolerance. The solution with the lowest final objective value across all restarts is retained. The MPS proxy is obtained by truncating the exact ground state to bond dimension $\chi$, with $\chi=1$ corresponding to a product-state approximation. The objective and its gradients are evaluated through tensor-network contractions, without explicitly enumerating all joint outcomes. After optimization, the resulting POVM is applied to the true state and its finite-shot RMSE is evaluated using the selected $k$-LO duals.

The fixed, non-optimized POVMs studied in Section~\ref{sec_rmse_non_opt} provide the baseline against which the optimized measurements are compared. In that section, we isolate the effect of the Platonic-solid geometry and the locality of the $k$-LO duals by keeping all POVM parameters fixed. Specifically, the same Platonic solid is used on every qubit, its canonical orientation is retained, and uniform probabilities are assigned to its $B=r/2$ antipodal vertex pairs. In terms of the parametrization introduced above, these choices correspond to
\begin{equation}
\begin{gathered}
    U^{(q)}=\id,
    \qquad
    \bm{\theta}^{(q)}=\bm{0},
    \qquad
    \nu_a^{(q)}=\frac{1}{B}=\frac{2}{r},\\
    a=1,\ldots,B.
\end{gathered}
\end{equation}
for every qubit $q=1,\ldots,n$. Thus, the measurements considered in Section~\ref{sec_rmse_non_opt} require no state or observable-dependent POVM optimization.

\section{Results for the non-optimized POVMs}\label{app_additional_results_non_optimized_povm}

In this Section, we present the numerical results for the RMSE for the \ce{H2} ($8$ qubits), \ce{LiH} ($12$ qubits)  and \ce{NH3} ($16$) molecules, which are also plotted in  the main text. The results are organized in tables that compare the RMSE obtained with different $k$-LO settings and the Platonic solid POVMs. Numerical results are shown in Tab.~\ref{tab_rmse_non_opt}. The best RMSE for each $k$-LO dual is highlighted in bold. The values for the different solids are also shown in Fig.~\ref{fig_rmse_non_opt}. The other molecules show a similar behaviour.

\begin{table*}[ht!]
\centering
\setlength{\tabcolsep}{8pt}
\renewcommand{\arraystretch}{1.2}
\begin{tabular}{c|c|c||cccc}
\toprule
\multirow{2}{*}{$k$-LO} &
\multirow{2}{*}{Molecule ($\#$ qubits)} &
\multirow{2}{*}{\shortstack{Uniform Pauli\\measurement~\cite{korhonen2025improvingshadowestimationlocallyoptimal}}} &
\multicolumn{4}{c}{Platonic-solid POVM}  \\
 & & & 
Octahedron & Cube & Icosahedron & Dodecahedron \\
\hline

\multirow{3}{*}{k=1} 
 & \ce{H2} (8)   & 0.060  & \textbf{0.0548} & 0.0578 & 0.0710 & 0.1404 \\
 & \ce{LiH} (12) & 0.032  & 0.0312 & \textbf{0.0310} & 0.0503 & 0.0526 \\
 & \ce{NH3} (16) & 0.353  & \textbf{0.3169} & 1.1969 & 0.9490 & 0.5888 \\
\hline

\multirow{3}{*}{k=2} 
 & \ce{H2} (8)    & 0.058 & 0.0558 & \textbf{0.0476} & 0.0486 & 0.0603 \\
 & \ce{LiH} (12)  & 0.029 & 0.0279 & \textbf{0.0267} & 0.0311 & 0.0502 \\
 & \ce{NH3} (16)  & \textbf{0.247} & 0.3194 & 0.4486 & 0.4893 & 0.4899 \\
\hline

\multirow{3}{*}{k=3} 
 & \ce{H2} (8)   & \multirow{3}{*}{--} & 0.0578 & \textbf{0.0488} & 0.0504 & 0.0640 \\
 & \ce{LiH} (12) &                     & 0.0275 & \textbf{0.0260} & 0.0272 & 0.0360 \\
 & \ce{NH3} (16) &                     & \textbf{0.2600} & 0.2661 & 0.4737 & 0.5996 \\
\hline

\multirow{3}{*}{k=4} 
 & \ce{H2} (8)   & 0.058  & 0.0581 & 0.0465 & \textbf{0.0415} & 0.0462 \\
 & \ce{LiH} (12) & \textbf{0.028} & 0.0281 & 0.0290 & 0.0387 & 0.0388 \\
 & \ce{NH3} (16) & \textbf{0.148} & 0.1612 & 0.2435 & 0.3395 & 0.3701 \\
\hline
\bottomrule
\end{tabular}

\caption{RMSE obtained with the fixed, non-optimized Platonic-solid POVMs for the different molecules and $k$-LO duals, compared with the uniform Pauli measurement reference values from Ref.~\cite{korhonen2025improvingshadowestimationlocallyoptimal}. The RMSE is computed using $R=10^3$ independent experimental realizations, each with $S=10^3$ measurement shots, while the $k$-LO duals are obtained from a separate dataset of $S=10^6$ shots. The best RMSE across the uniform Pauli measurement and the different Platonic-solid POVMs is highlighted in bold. The corresponding results are shown graphically in Fig.~\ref{fig_rmse_non_opt}.}

\label{tab_rmse_non_opt}
\end{table*}

\section{Results for the optimized POVMs on a proxy state}\label{app_additional_results_optimized_povm}

In this section, we present additional numerical results that complement the main findings of the paper. Namely, we report the RMSE values for the \ce{H2} ($8$ qubits) and \ce{BeH2} ($14$) molecules, which were not included in the main text for brevity. The results are organized in tables that compare the RMSE obtained with different bond dimensions $\chi$ of the MPS proxy state used in the optimization procedure, as well as the reference values from Ref.~\cite{korhonen2025improvingshadowestimationlocallyoptimal}. The same values of Tab.~\ref{tab_nh3_opt} for the \ce{NH3} molecule are also plotted in the main text. The best RMSE for each $k$-LO dual is highlighted in bold.

\begin{table*}[!ht]
\centering
\setlength{\tabcolsep}{8pt}
\renewcommand{\arraystretch}{1.2}

\begin{tabular}{c|c|c||ccc}
\toprule
\multirow{2}{*}{$k$-LO} &
\multirow{2}{*}{\shortstack{Platonic-solid\\POVM}} &
\multirow{2}{*}{\shortstack{Uniform Pauli\\measurement~\cite{korhonen2025improvingshadowestimationlocallyoptimal}}}
& \multicolumn{3}{c}{MPS proxy bond dimension $\chi$}  \\
 & & &
\normalfont{1} & \normalfont{5} & \normalfont{10} \\
\hline

\multirow{4}{*}{k=1} 
 & Octahedron   & \multirow{4}{*}{0.3530} & 2.9264 & 0.1959 & 0.1852 \\
 & Cube         &                         & 1.5831 & 0.6892 & 0.2074 \\
 & Icosahedron  &                         & 1.0850 & 0.2318 & \textbf{0.1835} \\
 & Dodecahedron &                         & 0.7786 & 0.2098 & 0.2391 \\
\hline

\multirow{4}{*}{k=2} 
 & Octahedron   & \multirow{4}{*}{0.2470} & 2.8635 & 0.1486 & 0.1558 \\
 & Cube         &                         & 0.6251 & 0.1474 & \textbf{0.1423} \\
 & Icosahedron  &                         & 0.4638 & 0.1580 & 0.1527 \\
 & Dodecahedron &                         & 1.1617 & 0.1653 & 0.1573 \\
\hline

\multirow{4}{*}{k=3} 
 & Octahedron   & \multirow{4}{*}{--} & 2.1089 & 0.1506 & \textbf{0.1319} \\
 & Cube         &                     & 0.5475 & 0.1382 & 0.1491 \\
 & Icosahedron  &                     & 0.4716 & 0.1420 & 0.1573 \\
 & Dodecahedron &                     & 1.0759 & 0.1499 & 0.1414 \\
\hline

\multirow{4}{*}{k=4} 
 & Octahedron   & \multirow{4}{*}{0.1480} & 0.7726 & 0.1491 & 0.1482 \\
 & Cube         &                         & 0.5807 & \textbf{0.1309} & 0.1361 \\
 & Icosahedron  &                         & 0.4900 & 0.1506 & 0.1451 \\
 & Dodecahedron &                         & 0.4044 & 0.1488 & 0.1586 \\
\hline
\bottomrule
\end{tabular}

\caption{RMSE for the \ce{NH3} molecule ($16$ qubits) obtained with optimized Platonic-solid POVMs and $k$-LO duals, compared with the uniform Pauli measurement reference values from Ref.~\cite{korhonen2025improvingshadowestimationlocallyoptimal}. The columns labeled by $\chi$ report results obtained by optimizing the POVM using an MPS proxy state with bond dimension $\chi=1$, $\chi=5$, or $\chi=10$. The RMSE is computed using $R=10^3$ independent experimental realizations, each with $S=10^3$ measurement shots, while the $k$-LO duals are obtained from a separate dataset of $S=10^6$ shots. The best RMSE for each $k$-LO dual is highlighted in bold.}

\label{tab_nh3_opt}
\end{table*}

\begin{table*}[!ht]
\centering
\setlength{\tabcolsep}{8pt}
\renewcommand{\arraystretch}{1.2}

\begin{tabular}{c|c|c||ccc}
\toprule
\multirow{2}{*}{$k$-LO} &
\multirow{2}{*}{\shortstack{Platonic-solid\\POVM}} &
\multirow{2}{*}{\shortstack{Uniform Pauli\\measurement~\cite{korhonen2025improvingshadowestimationlocallyoptimal}}}
& \multicolumn{3}{c}{MPS proxy bond dimension $\chi$}  \\
 & & &
\normalfont{1} & \normalfont{5} & \normalfont{10} \\
\hline

\multirow{4}{*}{1} 
 & Octahedron   & \multirow{4}{*}{0.1670} & 0.4555 & 0.0797 & 0.0819 \\
 & Cube         &                         & 0.2196 & 0.0956 & 0.0778 \\
 & Icosahedron  &                         & 1.3750 & 0.0871 & 0.0905 \\
 & Dodecahedron &                         & 0.3015 & 0.1324 & \textbf{0.0854} \\
\hline

\multirow{4}{*}{2} 
 & Octahedron   & \multirow{4}{*}{0.1510} & 0.1557 & \textbf{0.0654} & 0.0710 \\
 & Cube         &                         & 0.2966 & 0.0663 & 0.0706 \\
 & Icosahedron  &                         & 0.1059 & 0.0692 & 0.0741 \\
 & Dodecahedron &                         & 0.1273 & 0.0893 & 0.0730 \\
\hline

\multirow{4}{*}{3} 
 & Octahedron   & \multirow{4}{*}{--} & 0.1661 & \textbf{0.0648} & 0.0708 \\
 & Cube         &                     & 0.1124 & 0.0652 & 0.0705 \\
 & Icosahedron  &                     & 0.0999 & 0.0675 & 0.0724 \\
 & Dodecahedron &                     & 0.1160 & 0.0766 & 0.0716 \\
\hline

\multirow{4}{*}{4} 
 & Octahedron   & \multirow{4}{*}{0.1190} & 0.7284 & 0.0701 & 0.0704 \\
 & Cube         &                         & 0.1635 & 0.0687 & \textbf{0.0695} \\
 & Icosahedron  &                         & 0.0989 & 0.0722 & 0.0760 \\
 & Dodecahedron &                         & 0.1218 & 0.0962 & 0.0839 \\
\hline
\bottomrule
\end{tabular}

\caption{\ce{BeH2} (14 qubits) molecule. RMSE comparison between reference values and results for different bond dimensions. The best RMSE for each $k$-LO dual is highlighted in bold.}
\label{tab_beh2_opt}
\end{table*}

\begin{table*}[!ht]
\centering
\setlength{\tabcolsep}{8pt}
\renewcommand{\arraystretch}{1.2}

\begin{tabular}{c|c|c||ccc}
\toprule
\multirow{2}{*}{$k$-LO} &
\multirow{2}{*}{\shortstack{Platonic-solid\\POVM}} &
\multirow{2}{*}{\shortstack{Uniform Pauli\\measurement~\cite{korhonen2025improvingshadowestimationlocallyoptimal}}}
& \multicolumn{3}{c}{MPS proxy bond dimension $\chi$}  \\
 & & &
\normalfont{1} & \normalfont{5} & \normalfont{10} \\
\hline

\multirow{4}{*}{1} 
 & Octahedron   & \multirow{4}{*}{0.0600} & 0.0730 & 0.0480 & \textbf{0.0476} \\
 & Cube         &                         & 0.0866 & 0.0498 & 0.0486 \\
 & Icosahedron  &                         & 0.1082 & 0.0551 & 0.0529 \\
 & Dodecahedron &                         & 0.1139 & 0.0892 & 0.0505 \\
\hline

\multirow{4}{*}{2} 
 & Octahedron   & \multirow{4}{*}{0.0580} & 0.0578 & 0.0430 & \textbf{0.0438} \\
 & Cube         &                         & 0.0638 & 0.0447 & 0.0445 \\
 & Icosahedron  &                         & 0.0542 & 0.0492 & 0.0474 \\
 & Dodecahedron &                         & 0.0628 & 0.0851 & 0.0447 \\
\hline

\multirow{4}{*}{3} 
 & Octahedron   & \multirow{4}{*}{--} & 0.0635 & 0.0436 & \textbf{0.0421} \\
 & Cube         &                     & 0.0596 & 0.0430 & 0.0425 \\
 & Icosahedron  &                     & 0.0467 & 0.0474 & 0.0443 \\
 & Dodecahedron &                     & 0.0523 & 0.0779 & 0.0446 \\
\hline

\multirow{4}{*}{4} 
 & Octahedron   & \multirow{4}{*}{0.0580} & 0.0516 & 0.0428 & \textbf{0.0416} \\
 & Cube         &                         & 0.0473 & 0.0427 & 0.0438 \\
 & Icosahedron  &                         & 0.0503 & 0.0475 & 0.0448 \\
 & Dodecahedron &                         & 0.0465 & 0.0810 & 0.0445 \\
\hline
\bottomrule
\end{tabular}

\caption{\ce{H2} (8 qubits) molecule. RMSE comparison between reference values and results for different bond dimensions. The best RMSE for each $k$-LO dual is highlighted in bold.}
\label{tab_h2_opt}
\end{table*}

\begin{figure*}[!ht]
    \centering
    \includegraphics[width=\textwidth]{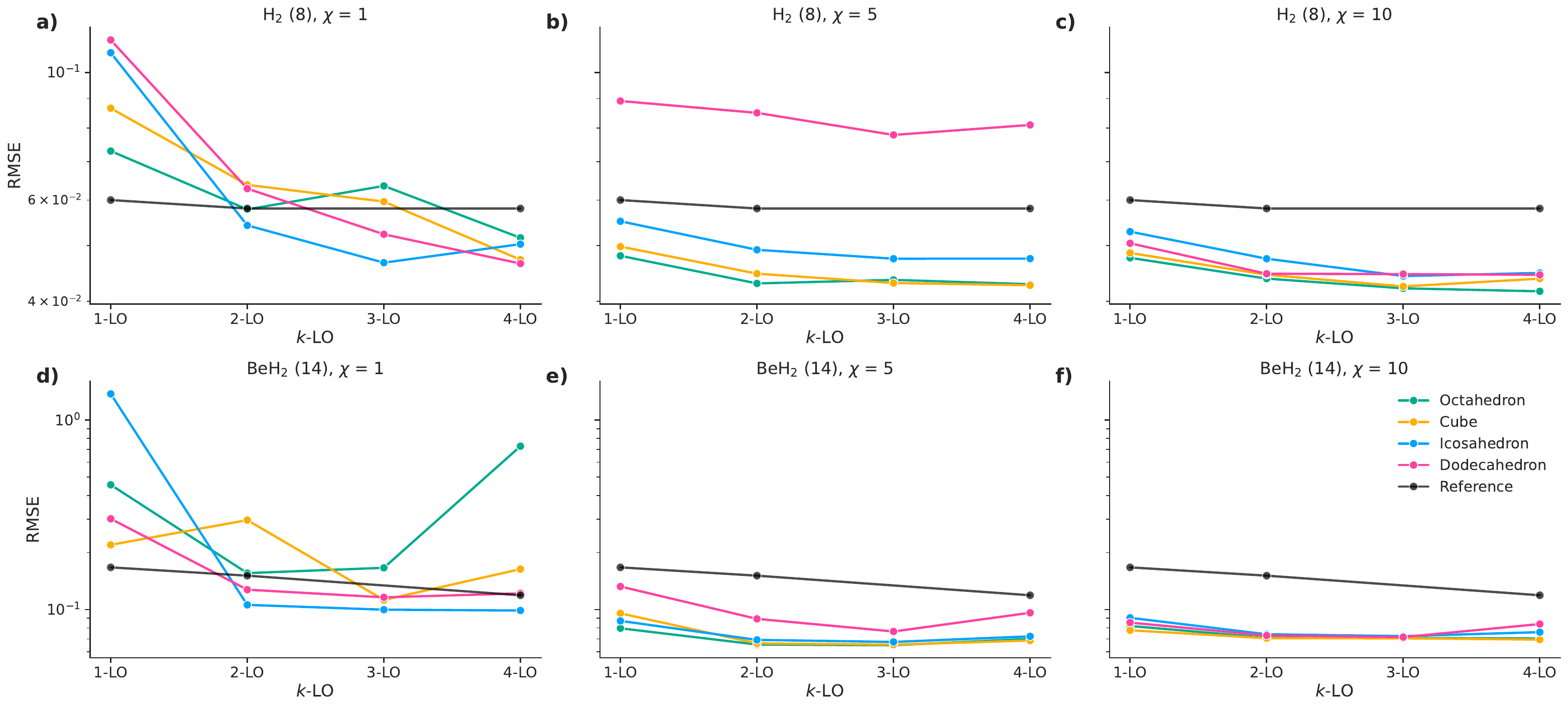}
    \caption{RMSE for the optimized Platonic-solid POVMs as a function of the locality parameter $k$ of the $k$-LO duals. The top row shows the results for \ce{H2} ($8$ qubits), while the bottom row shows the results for \ce{BeH2} ($14$ qubits). The three columns correspond to MPS proxy states with bond dimensions $\chi=1$, $\chi=5$, and $\chi=10$, respectively. The RMSE is computed using $R=10^3$ independent realizations, each with $S=10^3$ measurement shots, and $k$-LO duals obtained from a separate dataset of $S=10^6$ shots. Colored curves correspond to the different optimized Platonic-solid POVMs, while the black curve is the reference uniform Pauli POVM processed with the same $k$-LO duals, using the values reported in Ref.~\cite{korhonen2025improvingshadowestimationlocallyoptimal}. The corresponding numerical RMSE values are reported in Tabs.~\ref{tab_h2_opt} and~\ref{tab_beh2_opt}.}
    \label{fig_h2_beh2}
\end{figure*}

\section{Overfitting of the POVM optimisation}\label{app_overfit_analysis}

As discussed previously, the optimization procedure is performed using the MPS proxy state $\tilde{\rho}$ instead of the true state $\rho$. This means that the optimized POVM is not guaranteed to be optimal for the true state, and there is a risk of overfitting to the proxy state. 

\begin{figure*}[!ht]
    \centering
    \includegraphics[width=\textwidth]{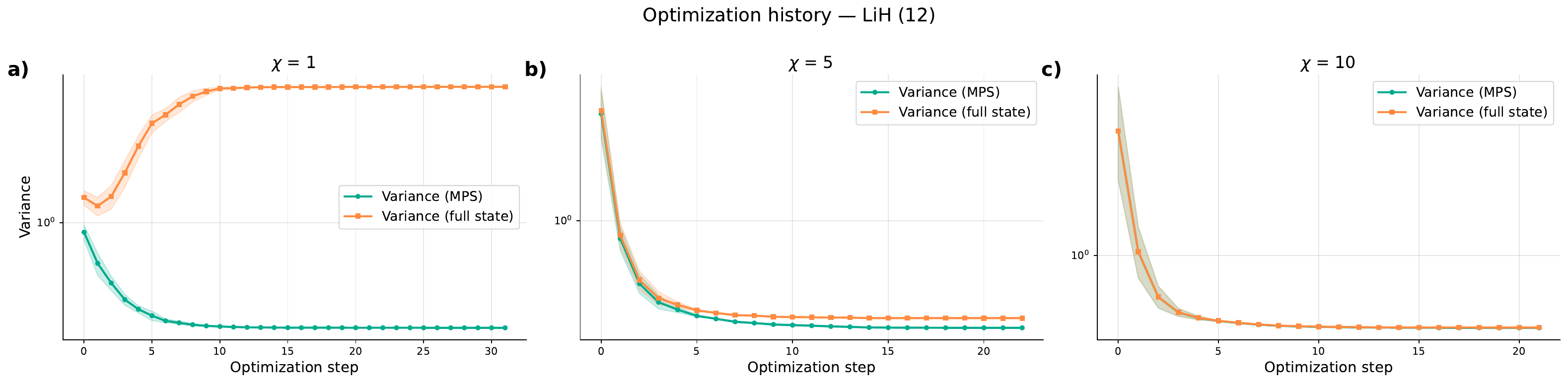}
    \caption{Variance comparison between the MPS proxy state $\tilde{\rho}$ and the true state $\rho$ during the optimization procedure for the \ce{LiH} (12 qubits) molecule, using the octahedron-based POVM. The variances are computed using the objective in Eq.~\eqref{eq_proxy_variance_objective} with $1$-LO duals, and the optimized POVM is obtained by minimizing $\Var{O}\big|_{\tilde{\rho}}$ over the POVM parameters. As the bond dimension $\chi$ increases, the variance obtained on the MPS proxy state becomes closer to the variance obtained on the true state. In contrast, for smaller bond dimensions, a larger gap is observed. Similar behaviour is observed for the other molecules and solids.}
    \label{fig_overfit}
\end{figure*}

To analyse this issue, we compare the variance $\Var{O}\big|_{\tilde{\rho}}$ obtained with the optimized POVM to the variance $\Var{O}\big|_{\rho}$ obtained with the same POVM evaluated on the true state during the optimization process. If the two variances are close, the optimized POVM is expected to generalize well to the true state. On the other hand, if there is a large gap between the two variances, this indicates that the optimization procedure is overfitting to the proxy state and does not perform well on the true state. Fig.~\ref{fig_overfit} shows this comparison for the octahedron-based POVM on the \ce{LiH} (12 qubits) molecule, across several values of the bond dimension $\chi$. As $\chi$ increases, the gap between the proxy variance and the true-state variance closes, confirming that a higher-fidelity MPS approximation leads to better generalization.

\end{document}